\documentclass[epj]{svjour}
\usepackage{graphics}
\usepackage{graphicx,epsfig}
\begin{document}
\title{
On the possibility to detect quantum correlation regions with the variable optimal
measurement angle
}

\author{Ekaterina~V.~Moreva\inst{1}\thanks{corresponding author: ekaterina.moreva@gmail.com},
Marco~Gramegna\inst{1}, \and
Mikhail~A.~Yurischev\inst{2}
}                     
%
%
\institute{Istituto Nazionale di Ricerca Metrologica, strada delle Cacce 91,
10135 Torino, Italy\and
Institute of Problems of Chemical Physics of the Russian Academy of Sciences,
142432 Chernogolovka, Moscow Region, Russia
}
\date{Revised version: date 14/01/2019}
%
\abstract{
Quantum correlations described by quantum discord and one-way quantum deficit can
contain ordinary regions with {\em constant} (i.e., universal) optimal measurement
angle $0$ or $\pi/2$ with respect to the $z$-axis and regions with a {\em variable}
(state-dependent) angle of the optimal measurement.
The latter regions which are absent in the Bell-diagonal
states are very tiny for the quantum discord and cannot be observed experimentally due to various
imperfections on the preparation and measurement steps of the experiment.
On the contrary, for the one-way quantum deficit we succeeded in getting the special two-qubit
X states which seem to allow one to reach all regions of quantum correlation exploiting available quantum optical techniques.
These states give possibility to deep investigation of quantum correlations and related optimization problems at new region and its boundaries.
In the paper, explicit theoretical calculations applicable to one-way deficit are reported,
together with the design of the experimental setup for generating such selected family of states; moreover, there are presented
numerical simulations showing that the most inaccessible region with the
intermediate optimal measurement angle may be resolved experimentally.}


\authorrunning{E.~V.~Moreva {\em et al.}}
\titlerunning{Quantum correlation regions with the variable optimal measurement angle}
\maketitle
%

\section{Introduction}
\label{intro}
Quantum correlations lie at the heart of quantum information science and technology.
Many kinds of quantum correlations have been introduced so far
and now their properties are scrupulously analyzed both theoretically and
experimentally.
Among quantumness quantifiers beyond quantum entanglement, relevant places in the scale of importance are occupied respectively by quantum discord and quantum work (information) deficit
\cite{MBCPV12,Str15,ABC16,FPA17,BDSRSS18}.

The quantum discord $Q$ for a bipartite system $AB$ is defined
as the minimum difference between the quantum generalizations of symmetric ($I$) and
asymmetric ($J$) forms of classical mutual information: $Q=\min_{\{{\rm\Pi_k}\}}(I-J)$,
where $\{{\rm\Pi_k}\}$ is the measurement performed on one of the two subsystems
\cite{Z00,OZ02} (see also \cite{HV01} in this regard).
The quantum discord is always non-negative, equals zero for the classically correlated
states, and coincides with the quantum entanglement for the pure states.
However, discord and entanglement exhibit essentially different behavior even for the
simplest mixed states --- the Werner and Bell-diagonal ones (see, e.g,
\cite{Y11,MGY17}).
Note that discord is not a symmetric quantity and in general it depends on which
subsystem the local measurement was performed.
A value of quantum discord for two-qubit systems can vary from zero to one bit.

The quantum work deficit is a measure of quantum correlation based on thermodynamics.
It was defined firstly by Oppenheim et al. \cite{OHHH02} as the difference between the work ${\cal W}$
which can be extracted from a heat bath using operations on the entire quantum system
and the largest amount of work $W$ drawn from the same heat bath by manipulating only
the local parts of composite system; in other words, the work deficit $\rm\Delta$ is the amount of potential work which
cannot be extracted under local operations and classical communication (LOCC) because of quantum correlations
\cite{OHHH02,HHHHOSS03,HHHOSSS05}.
Several forms of deficit exist, depending on the type of communication allowed
between parts $A$ (Alice) and $B$ (Bob).
For example, let us consider the case in which the bipartite state $\rho_{AB}$ is shared by Alice and Bob: if Bob
performs a single von Neumann measurement on his local subsystem and uses classical
communication to send the resulting state to Alice, when she extracts the maximum amount of
work $W$ from the new entire state, then the dimensionless quantity
${\rm\Delta}=\min_{\{{\rm\Pi_k}\}}({\cal W}-W)/k_BT$ is called the one-way quantum
deficit ($T$ is the temperature of the common bath and $k_B$ is Boltzmann's constant).
In spite of quite different conceptual sources, the one-way deficit and discord
coincide in considerably more general cases than entanglement.
They are the same for the Bell-diagonal states and even for the two-qubit X
states\footnote{
   A matrix having non-zero entries only along the diagonal and anti-diagonal
	 is called the X one because it looks like the letter ``X''.
}
with zero Bloch vector for one qubit if the local measurement is performed on this
qubit \cite{YF16}.
On the other hand, these quantum correlations exhibit, generally speaking, different
quantitative and qualitative behavior in more general cases.

Due to the optimization procedure entering in quantum correlation definitions,
evaluation of quantum discord and deficit entails great difficulties even for the
two-qubit systems.
Up to the present, closed analytical formula for the quantum discord has been derived
only for a particular class of X states, namely for the
Bell-diagonal states~\cite{Luo08}.
Since the one-way deficit is identical to the discord in this case, one automatically
possesses the closed analytical expression for the former quantity.

An attempt to extend the success of Luo \cite{Luo08} to the arbitrary X states was
undertaken in 2010 by Ali, Rau, and Alber \cite{ARA10}.
Unfortunately, the authors decided that the extreme values of parameters which
characterize the von Neumann measurement were attained for discord only at their
endpoints.
Shortly after, however, the counterexamples of X density matrices have been
given which demonstrate a measurement-dependent discord minimum inside
the interval of measurement parameters \cite{LMXW11,CZYYO11}.
Thus, the analytic formula of Ref.~\cite{ARA10} is incorrect in general.

At that time it was also established that for the general two-qubit X states the
optimization of discord over the projectors $\{{\rm\Pi}_k(\theta,\varphi)\}$ can be
worked out exactly over the azimuthal angle $\varphi$ but one optimization procedure,
in the polar angle $\theta\in[0,\pi/2]$, remains relevant \cite{LXSW10,CRC10,VR12}
(see also \cite{JY16}).
As a result, a pessimistic verdict has been made:
``For general two-qubit X states quantum discord cannot be evaluated
analytically'' \cite{H13}.

Definite optimism was restored in Refs.~\cite{Y14,Y14a,Y15}, where
it has been observed that the formula for calculating the quantum discord
of general two-qubit X states has, in any event, a piecewise-analytical-numerical
(semianalytical) form
\begin{equation}
   \label{eq:Q3}
   Q=\min\{Q_0, Q_{\theta^*}, Q_{\pi/2}\}.
\end{equation}
Here the subfunctions (branches) $Q_0$ and $Q_{\pi/2}$ are the analytical expressions
(corresponding to the discord with optimal measurement angles 0 and $\pi/2$,
respectively) and only the third branch $Q_{\theta^*}$ requires one-dimensional
searching of the optimal state-dependent measurement angle $\theta^*\in(0,\pi/2)$ if,
of course, the interior global extremum exists.

Thus, the total domain of definition for the discord function consists of subdomains
each one corresponding to the own branch (phase or fraction - in physical language)
separated by strong boundaries.
Equations for such boundaries have been proposed in Refs.~\cite{Y14,Y14a,Y15}.
The equations for 0- and $\pi/2$-boundaries separating respectively the $Q_0$ and
$Q_{\pi/2}$ regions with the $Q_{\theta^*}$ one are written as
\begin{equation}
   \label{eq:QII}
   Q^{\prime\prime}(0)=0, \qquad   Q^{\prime\prime}(\pi/2)=0.
\end{equation}
Here $Q^{\prime\prime}(0)$ and $Q^{\prime\prime}(\pi/2)$ are the second derivatives
of the measurement-dependent discord function $Q(\theta)$ with respect to  $\theta$
at the endpoints $\theta=0$ and $\pi/2$, correspondingly.
The equations (\ref{eq:QII}) are based on the unimodality hypothesis for the function
$Q(\theta)$ and bifurcation mechanism of appearance of the extremum inside the
interval $(0,\pi/2)$.
The equations are confirmed now for different subclasses of X states \cite{Y15,Y17}.

Very similar situation takes place for the one-way quantum deficit of two-qubit X
states \cite{YWF16,Y18,Y18a}.
This quantity is given as
\begin{equation}
   \label{eq:D3}
   {\rm\Delta}=\min\{\Delta_0, \Delta_{\vartheta}, \Delta_{\pi/2}\},
\end{equation}
where the branches $\Delta_0$ and $\Delta_{\pi/2}$ are again known in the analytical
form while the third branch $\Delta_{\vartheta}$ requires to perform numerical
minimization to obtain state-dependent minimizing polar angle $\vartheta\in(0,\pi/2)$.
However the measurement-dependent deficit $\Delta(\theta)$ can exhibit now the bimodal
behavior that in turns can lead additionally to the new mechanism of formation
of a boundary between the phases, namely via finite jumps of optimal measured angle
from the endpoint to the interior minimum or vice versa \cite{Y18,Y18a}.

The analysis performed shows that the discordant region $Q_{\theta^*}$ is very narrow.
It is characterized by the linear sizes of order $10^{-4}$, leading to the
fantastically high fidelity\footnote{
   The fidelity of two quantum states, $F$, leads to the Bures distance, $d_B$,
	 between the same states through the relation $d_B=[2(1-\sqrt{F})]^{1/2}$.
}
between the boundary states: $F=99.999998\%$ \cite{Y17}.
Moreover, the volume of $Q_{\theta^*}$-region is $0.08\%$ of total volume of the
domain of definition \cite{Y17}.
The latter agrees approximately with the  estimation obtained by Monte-Carlo
simulations \cite{CDSPSS15} equivalent to $0.03\%$.
Thus, these parameters are unfavorable and exclude any possibility to observe now the
exotic $Q_{\theta^*}$-region experimentally.

On the other hand, the analogous regions $\Delta_{\vartheta}$ of one-way quantum
deficit can achieve the sizes comparable to those of the regions $\Delta_0$ and
$\Delta_{\pi/2}$, therefore inspiring the possibility that an insight into the
considered region can be obtained experimentally.

The aim of present paper is to select suitable quantum states showing the widest
possible regions with the variable intermediate optimal measurement angle, to perform for them
numerical simulations, and to give a response about the possibility to resolve such
regions using contemporary optical apparatus.

Our aspiration to experimentally detect the new regions (phases) of quantum
correlations is motivated by the following.
First, this is a study of properties of quantum correlations which are absent in the Bell-diagonal states.
In particular, the observation of continuous and smooth transitions between the phases which manifest in 
higher derivatives with respect to the state parameters.
Secondly, the fact that the state lies in the region of the variable optimal measurement allows us 
to estimate the value of quantum correlation via the shift of the angle.

Third, the regions of quantum states with the intermediate optimal measurement angle are rather 
surprising because the most practical constrained optimization problems in the natural sciences 
have an optimal solution at the boundary.(See, e.g., \cite{RSIS09}:
``Real life optimization problems often involves one or more constraints and in most
cases, the optimal solutions to such problems lie on constraint boundaries.'')
However this expectation can lead to incorrect results like in \cite{ARA10}.

In the following sections, a suitable candidate of quantum state is given, its
properties are described in detail, a scheme of optical setup is considered and
the expected results are discussed.
Finally, in the last section, a brief conclusion is given.

\section{
Theoretical results
}
\label{sect:Theory}
We begin with the theoretical description of the problem under question.

\subsection{One-way deficit
 estimation}
\label{sec:defDef}
The maximum amount of useful work that can be extracted from a
system in the state $\rho$ is given as
\cite{OHHH02,HHHHOSS03,HHHOSSS05} (see also \cite{MBCPV12,Str15,ABC16})
\begin{equation}
   \label{eq:w}
   w=k_BT(\log d - S(\rho)),
\end{equation}
where $S(\rho)=-{\rm tr}(\rho\log\rho)$ is the entropy of state $\rho$ and $d$ the
dimension of Hilbert space in which the density operator $\rho$ acts.
Applying this general relation to the states before and after Bob's measurement it is possible to obtain the following equation for the one-way deficit
\begin{equation}
   \label{eq:rmD}
   {\rm\Delta}=\min_{\{\rm\Pi_k\}}S(\tilde\rho_{AB})-S(\rho_{AB}),
\end{equation}
where
\begin{equation}
   \label{eq:rho_tilde}
   \tilde\rho_{AB}\equiv\sum_kp_k\rho^k_{AB}
	 =\sum_k({\rm I}\otimes{\rm\Pi}_k)\rho_{AB}({\rm I}\otimes{\rm\Pi}_k)^+
\end{equation}
is the weighted average of post-measured states
\begin{equation}
   \label{eq:rho-k}
   \rho^k_{AB}
	 =\frac{1}{p_k}({\rm I}\otimes{\rm\Pi}_k)\rho_{AB}({\rm I}\otimes{\rm\Pi}_k)^+
\end{equation}
with the probabilities
\begin{equation}
   \label{eq:p_k}
	 p_k={\rm Tr}({\rm I}\otimes{\rm\Pi}_k)\rho_{AB}({\rm I}\otimes{\rm\Pi}_k)^+.
\end{equation}
Thus, the one-way quantum deficit equals the minimal increase of entropy after a von
Neumann measurement on one party of the bipartite system $\rho_{AB}$.

It is clear from Eq.~(\ref{eq:rmD}) that the main problem is to find the
post-measurement entropy, because the pre-measurement one, i.e.  $S(\rho_{AB})$, does not
depend on the measuring angle and hence plays a role of a trivial constant shift.
Therefore, below we will stress attention mainly on the $S(\tilde\rho_{AB})$.

Generally the one-way deficit, as the quantum discord, is asymmetric quantity under
replacement of the measured subsystem, however we avoid such cases in our work.

Notice that the map $\rho_{AB}\mapsto\tilde\rho_{AB}$ defined by
Eq.~(\ref{eq:rho_tilde}) can be interpreted as non-selective measurement
(see, e.g., the textbook \cite{I12}) because not the individual measurement outcomes
are recorded but only the statistics of outcomes is known.
(Note in passing that the quantum discord is based on selective measurements.)
Moreover, this map has a form of quantum operation and therefore
the one-way deficit has the operational significance beyond entanglement.

It arises a question: how to determine the entropy of some quantum state $\rho$
experimentally?
In order to get the thermodynamic entropy one measures the heat capacity of the given
sample, takes the ratio of heat capacity to the temperature and then integrates this
ratio with respect to the temperature.
In the quantum case, in line with the measurement propositions
\cite{D30,vN32}, direct way is to take the entropy operator $-\log\rho$ \cite{Z71}
or the density one $\rho$ as an observable.
However, it is not known how to experimentally realize the projectors
$|\lambda_k\rangle\langle\lambda_k|$, where $|\lambda_k\rangle$ are the eigenvectors
of above operators.
Instead, one can first restore the quantum state $\rho$ through a
tomographic reconstruction in the computational basis (i.e., find the numerical values
for all entries of the density matrix), solve eigenvalue problem for this matrix on a
computer, and then calculate the quantum entropy via the relation
$S(\rho)=-\sum_i\lambda_i\log{\lambda_i}$ with $\lambda_i$
being the eigenvalues of $\rho$ \cite{JKMW01}.
It is the way that we use in the present paper.

There is, of course, the radical way to obtain the one-way deficit without performing
any local measurements at all (as it is usually made to get the quantum entanglement;
say, \cite{BSPBG13}).
Indeed, since the full tomography is needed to find the entropy of pre-measurement
state, one can use the digital representation of $\rho_{AB}$ to numerically perform
the required local measurement, compute the minimized post-measurement entropy, and
finally arrive at the value of one-way deficit $\rm\Delta$.
We keep in mind this possibility and will compare both approaches in the real
experimental work.
Both possibilities have their pluses and minuses.
If the local measurement is performed in analog way, then the numerical calculations
are simplified, moreover we prefer to consider "real" measurements with their imperfections for the simulation of the experiment.

To continue our consideration one should specify the quantum state.

\subsection{
Initial quantum state $\rho_{AB}$
}
\label{sec:rho_AB}
Focussing on two-qubit systems,
to this date, phase diagrams for the quantum discord and one-way deficit have been
studied in detail for the three-parameter subclass of two-qubit X states
\cite{Y17,Y18a}.
This allows us to choose the suitable state to examine it in an experiment.

From the available variety of states, we consider here the maximally simple (but
non-trivial) one-parameter state
\begin{equation}
   \label{eq:rho}
   \rho_{AB}=q|\Phi^+\rangle\langle\Phi^+|
	 + (1-q)|01\rangle\langle01|,
\end{equation}
where $|\Phi^+\rangle=(|00\rangle+|11\rangle)/\sqrt{2}$.
This state in a Bloch form is written as
\begin{eqnarray}
   \label{eq:rho-Bloch}
   \rho_{AB}&&=4^{-1}[{\rm I}\otimes{\rm I}
   + (1-q)(\sigma_z\otimes{\rm I}
   - {\rm I}\otimes\sigma_z)
	 \nonumber\\
   &&
   + q(\sigma_x\otimes\sigma_x
   - \sigma_y\otimes\sigma_y)
   + (2q-1)\sigma_z\otimes\sigma_z],
\end{eqnarray}
where $\sigma_\alpha$ ($\alpha=x,y,z$) is the vector of the Pauli matrices.
The given state will show the obvious symmetry under permutations of particles
($A\leftrightarrow B$) after performing the local unitary (orthogonal) transformation
$U={\rm I}\otimes\sigma_x$ which does not change any of the quantum correlations.
Lastly, the density matrix of chosen state $\rho_{AB}$ in explicit form is
given by
\begin{equation}
   \label{eq:rho1}
   \rho_{AB}
	 =\left(
      \begin{array}{cccc}
      q/2&0&0&q/2\\
      0&1-q&0&0\\
      0&0&0&0\\
      q/2&0&0&q/2\\
      \end{array}
   \right).
\end{equation}

Let us consider now the class of states (\ref{eq:rho}) showing the maximum amount of entanglement
investigated in Refs.~\cite{PABJWK04,BMNM04,APVW07}.
It is remarkable that the authors~\cite{PABJWK04} were able to achieve the fidelity
$F\geq99\%$.
Later the discordant features of the above state were discussed in Ref.~\cite{GGZ11}.
Notice that the quantum discord for the state (\ref{eq:rho}) is defined by the optimal
measurement angle $\theta^*=\pi/2$ in the whole interval of parameter $q$ and
hence the discord has here no regions with the variable optimal measurement angles.

In this paper we focus on the one-way deficit.
According to Eq.~(\ref{eq:rmD}), to find this quantity, one should first
calculate the pre- and post-measurement entropies --- $S(\rho_{AB})$ and
$S(\tilde\rho_{AB})$, respectively.
For this purpose, we find the corresponding eigenvalues.

Eigenvalues of matrix (\ref{eq:rho1}) equal
\begin{equation}
   \label{eq:lam}
   \lambda_1=q,\quad \lambda_2=1-q,\quad \lambda_3=\lambda_4=0.
\end{equation}
Owing to the non-negativity requirement for any density matrix, one obtains that the
domain of definition for the parameter (argument) $q$ is restricted by the condition
\begin{equation}
   \label{eq:q1q2}
   0\leq q\le1.
\end{equation}
The quantity $q$ may be interpreted as a concentration of Bell-diagonal state in the
two-component mixture (\ref{eq:rho}).

Using Eq.~(\ref{eq:lam}) one gets the pre-measured entropy function
\begin{equation}
   \label{eq:preS}
   S(q)\equiv S(\rho_{AB})=-q\log{q}-(1-q)\log{(1-q)}.
\end{equation}
This is exactly the binary entropy and its value can vary from zero to one bit.

\subsection{
Post-measurement state $\tilde\rho_{AB}$
}
\label{sec:tildeS}
Since $\rho_{AB}$ is the two-qubit state, then $\rm\Pi_k$ in Eq.~(\ref{eq:rho_tilde})
are the two projectors  ($k=0,1$)
\begin{equation}
   \label{eq:Pi}
   {\rm\Pi}_k=V\pi_kV^+,
\end{equation}
where $\pi_k=|k\rangle\langle k|$ and transformations $\{V\}$ belong to
the special unitary group $SU_2$.
Rotations $V$ are parametrized by two angles $\theta$ and $\varphi$
(polar and azimuthal, respectively):
\begin{equation}
   \label{eq:V}
   V
	 =\left(
      \begin{array}{cc}
      \cos(\theta/2)&e^{i\varphi}\sin(\theta/2)\\
      \sin(\theta/2)&-e^{i\varphi}\cos(\theta/2)
      \end{array}
   \right)
\end{equation}
with $0\le\theta\le\pi$ and $0\le\varphi<2\pi$.

Performing the necessary calculations it is possible to get the eigenvalues of the density matrix
${\tilde\rho}_{AB}$:
\begin{eqnarray}
   \label{eq:Lam}
	 \Lambda_{1,2}&&=\frac{1}{4}\lbrack\!\lbrack1+(1-q)\cos\theta\pm\{[1-q+(1-2q)\cos\theta]^2
	 \nonumber\\
	 &&+q^2\sin^2\theta\}^{1/2}\rbrack\!\rbrack
	 \nonumber\\
	 \\
	 \Lambda_{3,4}&&=\frac{1}{4}\lbrack\!\lbrack1-(1-q)\cos\theta\pm\{[1-q-(1-2q)\cos\theta]^2
	 \nonumber\\
	 &&+q^2\sin^2\theta\}^{1/2}\rbrack\!\rbrack.
	 \nonumber
\end{eqnarray}
It is seen that the azimuthal angle $\varphi$ has dropped out from the given expressions.
This is due to the fact that one pair of anti-diagonal entries of the density matrix
(\ref{eq:rho1}) vanishes.
Using Eqs.~(\ref{eq:Lam}) we arrive at the post-measured entropy (entropy after
measurement)
\begin{equation}
   \label{eq:postS}
   \tilde S(\theta;q)\equiv S(\tilde\rho_{AB})=h_4(\Lambda_1,\Lambda_2,\Lambda_3,\Lambda_4),
\end{equation}
where $h_4(x_1,x_2,x_3,x_4)=-\sum_{i=1}^4x_i\log x_i$, with the additional condition
$x_1+x_2+x_3+x_4=1$, is the quaternary entropy function.
Notice that function $\tilde S$ of argument $\theta$ is invariant under
the transformation $\theta\to\pi-\theta$ therefore it is enough to consider the values for which $\theta\in[0,\pi/2]$.

It is worth noticing that since the post-measurement state is needed only
to find its entropy, which is invariant under any unitary transformations, one can
transfer the rotations $V$ from the measurement operators (projectors) on the state
$\rho_{AB}$:
\begin{eqnarray}
   \label{eq:rho_tilde1}
\tilde\rho_{AB}&&\mapsto\tilde\rho_{AB}^{\,\prime}
=({\rm I}\otimes\pi_0)\cdot[({\rm I}\otimes V^+)\rho_{AB}({\rm I}\otimes V)]\cdot({\rm I}\otimes\pi_0)
	 \nonumber\\
&&+({\rm I}\otimes\pi_1)\cdot[({\rm I}\otimes V^+)\rho_{AB}({\rm I}\otimes V)]\cdot({\rm I}\otimes\pi_1)
\end{eqnarray}
and in this case $S(\tilde\rho_{AB})=S(\tilde\rho_{AB}^{\,\prime})$.
It means that we may firstly rotate the state (e.g., with a half-wave plate)
and then perform two orthogonal projections of the rotated state in the {\em initial}
computational basis.

\subsection{
Behavior of post-measurement entropy
}
\label{sec:tildeD}
We describe here specific properties of post-measurement entropy which will be needed
for performing the experiment.
In other words, we shall try to supply experimentalists with technological maps suitable
in the work.

\begin{figure}[t]
\begin{center}
\epsfig{file=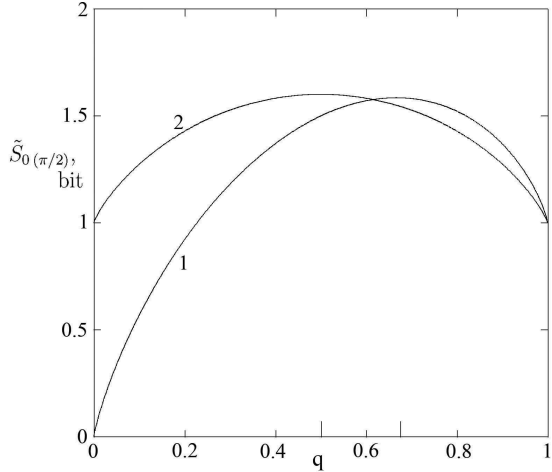,width=7.2cm}
\caption{
Dependencies $\tilde S_0(q)$ (curve 1) and $\tilde S_{\pi/2}(q)$ (curve 2).
Longer bars mark the position of interval $[0.5,0.67515]$
}
\label{fig:zsp01m}       
\end{center}
\end{figure}
Equations~(\ref{eq:preS}), (\ref{eq:Lam}), and (\ref{eq:postS}) define the
measurement-dependent (non-optimized) one-way deficit function
$\Delta(\theta)={\tilde S}(\theta)-S$.
Direct calculations show that for every choice of model parameter $q$ the function
${\tilde S}(\theta)$ and hence $\Delta(\theta)$ possess an important property,
namely their first derivatives with respect to $\theta$ identically equal to zero at
both endpoints $\theta=0$ and $\theta=\pi/2$:
\begin{equation}
   \label{eq:postSD1}
   {\tilde S}^{\prime}(0)=\Delta^{\prime}(0)\equiv0,\qquad
   {\tilde S}^{\prime}(\pi/2)=\Delta^{\prime}(\pi/2)\equiv0.
\end{equation}

\begin{figure}[t]
\begin{center}
\epsfig{file=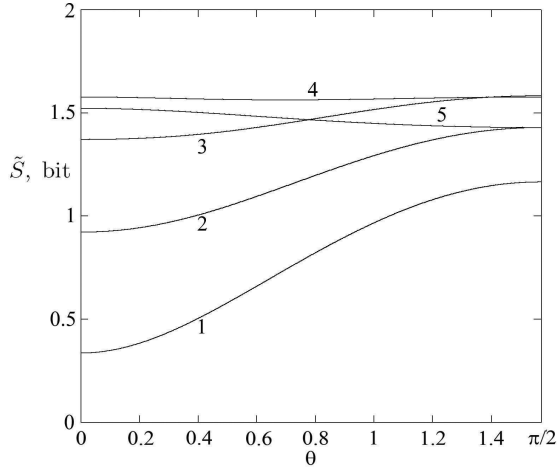,width=7.2cm}
\caption{
Post-measurement entropy $\tilde S$ vs $\theta$ by $q=0.05$~(1),
0.2~(2), 0.4~(3), 0.6155~(4), and 0.8~(5)
}
\label{fig:zs0005-08}       
\end{center}
\end{figure}
From Eqs.~(\ref{eq:Lam}) and (\ref{eq:postS}) we get the expressions for the
post-measurement entropy at the endpoint $\theta=0$,
\begin{equation}
   \label{eq:S0}
   \tilde S_0(q)=-(1-q)\log(1-q)-q\log(q/2),
\end{equation}
and at the second endpoint $\theta=\pi/2$:
\begin{equation}
   \label{eq:S1}
   \tilde S_{\pi/2}(q)=\log2+h((1+\sqrt{(1-q)^2+q^2})/2),
\end{equation}
where $h(x)=-x\log x-(1-x)\log(1-x)$ is the Shannon binary entropy function.
The behavior of functions $\tilde S_0(q)$ and $\tilde S_{\pi/2}(q)$ is depicted in
Fig.~\ref{fig:zsp01m}.
The local maxima of these functions lie at $q=2/3$ and $1/2$, and equal
$\rm log_23\approx1.58496$~bits and
$\frac{1}{4}[10-\sqrt2\,{\rm log_2}(3+2\sqrt2)]\approx1.60088$~bits, respectively.
The function $\tilde S_{\pi/2}(q)$ is symmetric under the replacement $q\to1-q$.
The curves 1 and 2 intersect at the point $(0.61554,1.57667)$.

Together with Eq.~(\ref{eq:preS}), the relations (\ref{eq:S0}) and (\ref{eq:S1}) supply
us with explicit expressions for the one-way deficit at the endpoints:
$\Delta_0=\Delta(0)$ and $\Delta_{\pi/2}=\Delta(\pi/2)$.
In particular, $\Delta_0=q\log2~(=q~{\rm bits})$.

At the 0- and $\pi/2$-boundaries, the second derivatives of the deficit and the post-measurement entropy are:
\begin{equation}
   \label{eq:D11}
   \Delta^{\prime\prime}(0)=0\quad
   {\rm and}\quad
   \Delta^{\prime\prime}(\pi/2)=0
\end{equation}

\begin{equation}
   \label{eq:S11}
   \tilde S^{\prime\prime}(0)=0\quad
   {\rm and}\quad
   \tilde S^{\prime\prime}(\pi/2)=0
\end{equation}
will be needed below.

As calculations yield,
\begin{equation}
   \label{eq:S110}
   \tilde S^{\prime\prime}(0)=\frac{1-3q+2q^2}{2-3q}\ln\frac{2(1-q)}{q}.
\end{equation}
The roots of equation $\tilde S^{\prime\prime}(0)=0$ are 1/2 and 1.

On the other hand, calculations show that the second derivative
$ \tilde S^{\prime\prime}(\theta)$ with respect to $\theta$ at $\theta=\pi/2$
equals
\begin{eqnarray}
   \label{eq:S11P}
	 \tilde S^{\prime\prime}(\pi/2)&&=\frac{q^2}{2r^3}[r^2-(1-2q)^2]\ln\frac{1+r}{1-r}
	 \nonumber\\
	 &&-\frac{(1-q)^2}{1-r^2}[1-2(1-2q)(1-\frac{1-2q}{2r^2})],
\end{eqnarray}
where
\begin{equation}
   \label{eq:r}
   r=\sqrt{(1-q)^2+q^2}.
\end{equation}
The results of numerical solution of the equation $\tilde S^{\prime\prime}(\pi/2)=0$
are $q=0.67515$ and again an uninteresting root equaled 1.
Thus, the region with the interior optimal measurement angle can exist only when
$q\in(0.5,0.67515)$.

%
\begin{figure*}[t]
\begin{center}
\epsfig{file=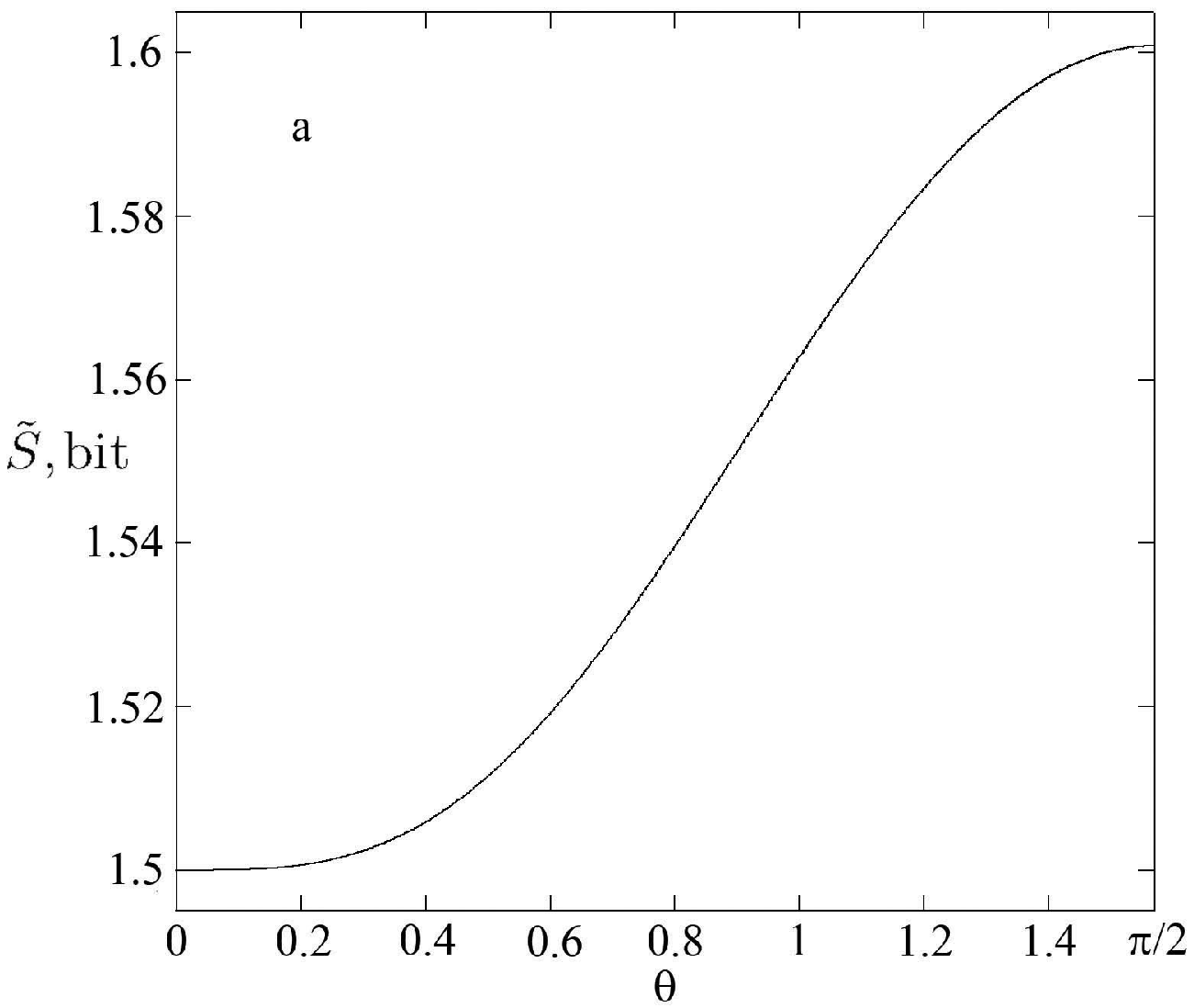,width=4.4cm}
\epsfig{file=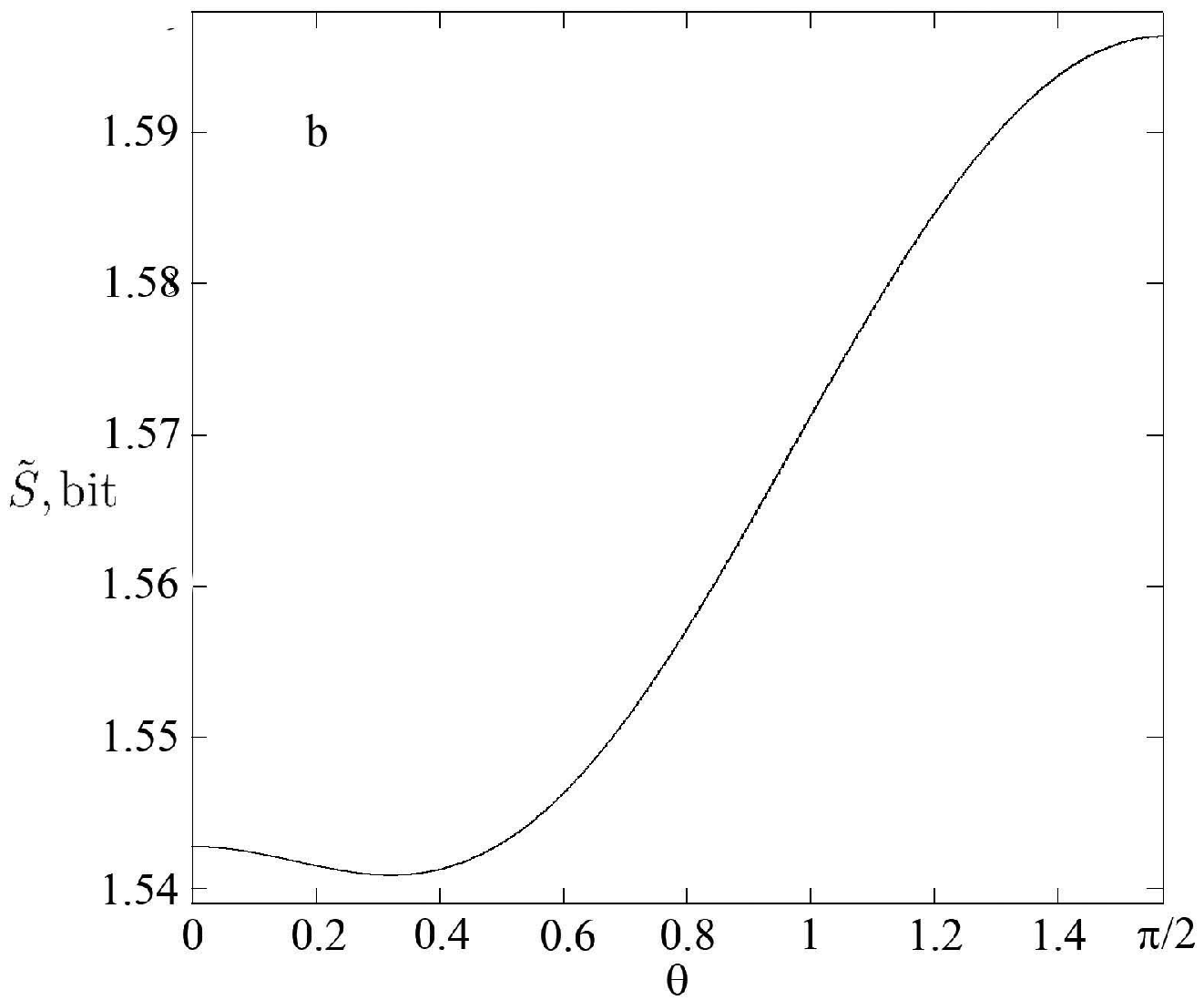,width=4.4cm}
\epsfig{file=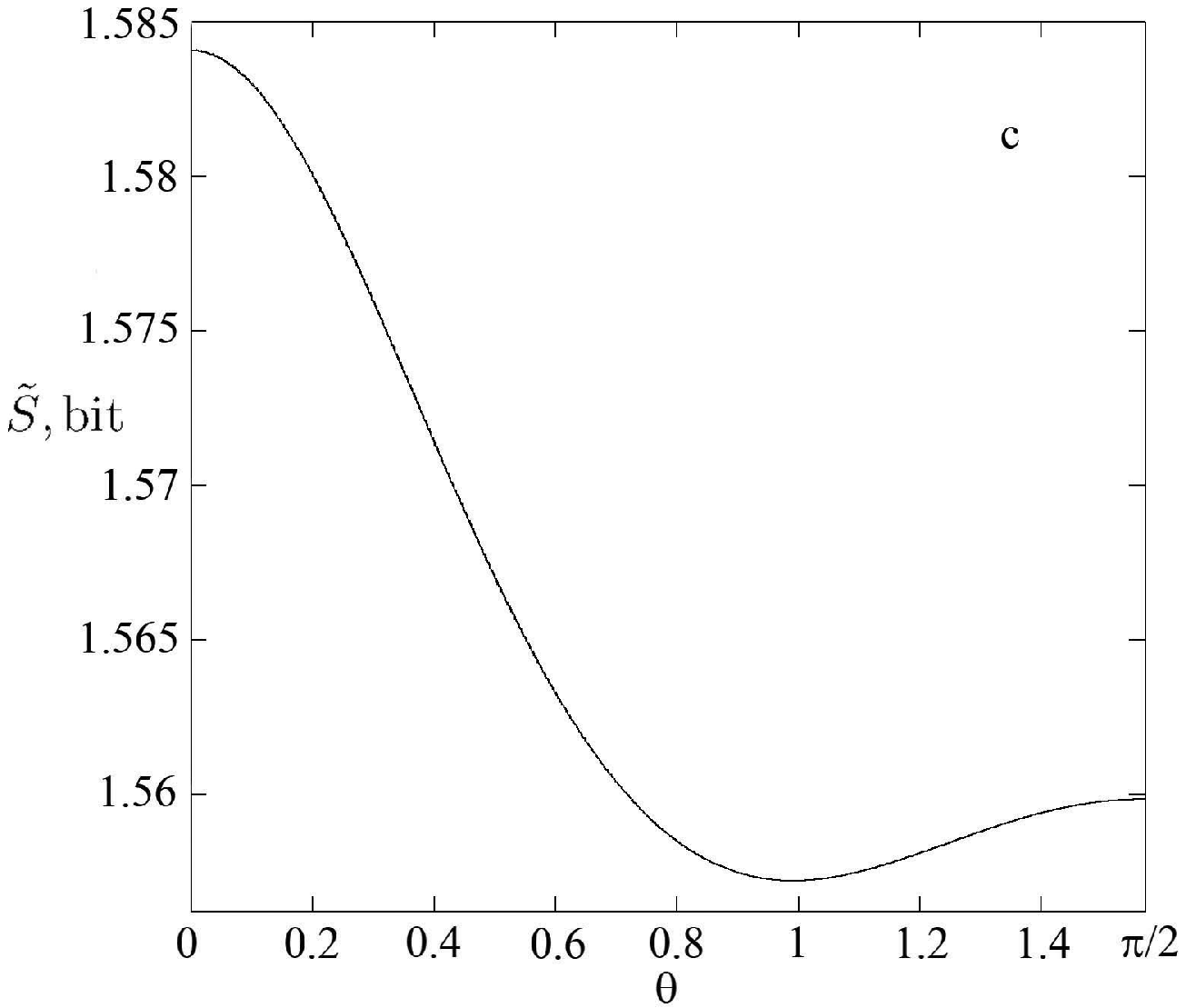,width=4.4cm}
\epsfig{file=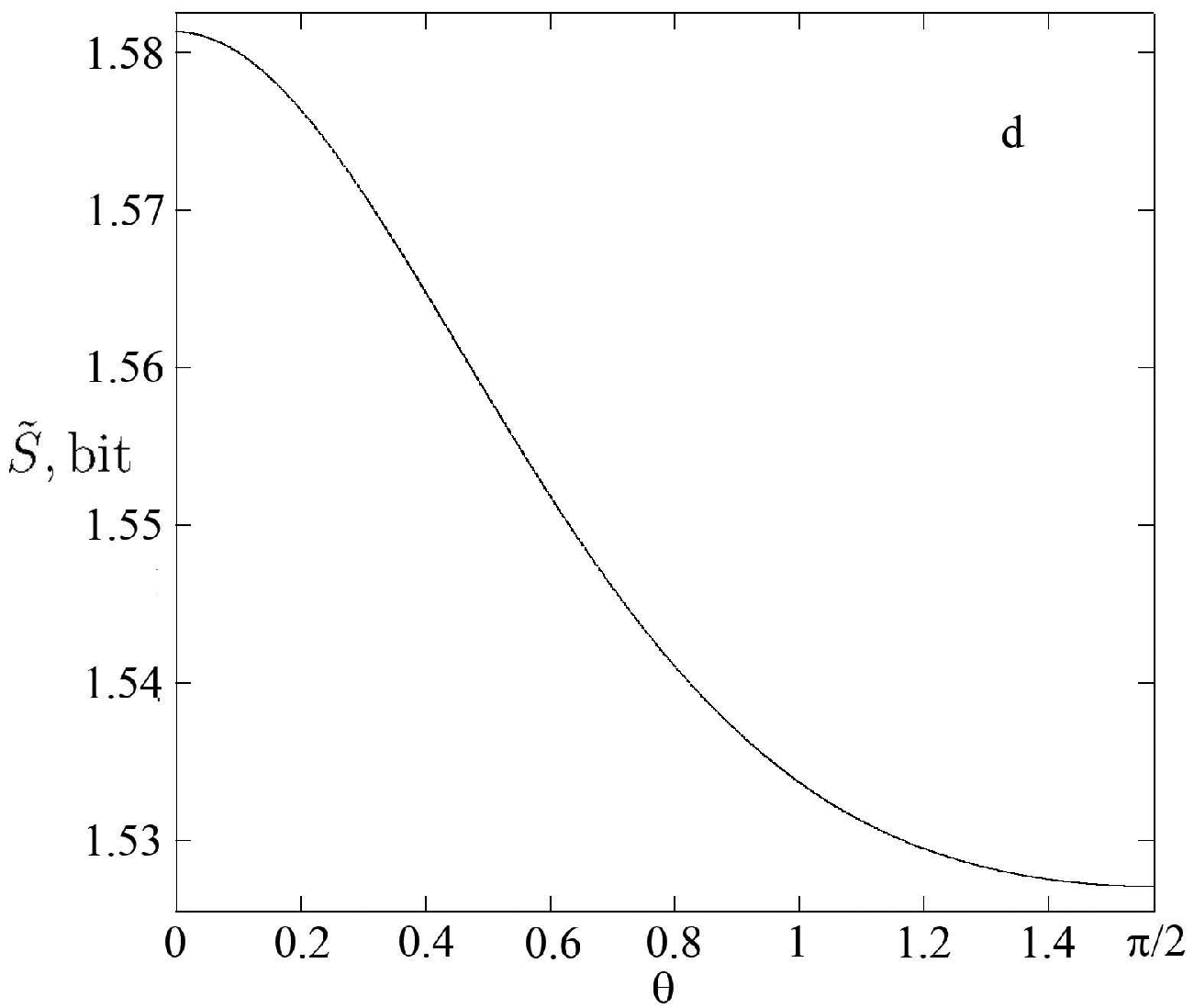,width=4.4cm}
\caption{
Post-measurement entropy $\tilde S$ vs $\theta$ by $q=0.5$~(a),
0.55~(b), 0.65~(c), and 0.7~(d)
}
\label{fig:zs4}       
\end{center}
\end{figure*}
%
Let us consider now the behavior of post-measured entropy for different values of
parameter $q$ in the segment $[0,1]$.
The values of entropy in two-qubit systems can vary from zero to two bits.
Figure~\ref{fig:zs0005-08} shows the the behavior of post-measurement entropy upon the
measurement angle $\theta$ by different values of parameter $q$.
The curves $\tilde S(\theta)$ for $q\le0.5$ have the monotonically increasing behavior
and here the optimal measurement angle is constant equaling zero.
The relative difference between values of entropy at endpoints $\theta=0$ and $\pi/2$
is large and achieves 71\%, 35\%, and 13\% respectively for $q=0.05$, 0.2, and 0.4
(see Fig.~\ref{fig:zs0005-08}).
Conversely, the curves of post-measurement entropy exhibit the monotonically
decreasing dependence for $q>0.67515$.
For example, the relative difference $(\tilde S(0)-\tilde S(\pi/2))/\tilde S(0)$
equals 6\% for $q=0.8$ (see again Fig.~\ref{fig:zs0005-08}).
So, the presented estimates are large enough and therefore allow to hope that the
effect of correlation-function transition from the optimal measurement angle zero to
$\pi/2$ can be observed in an experiment.

Let us discuss now the behavior of post-measurement entropy in the intermediate region
$0.5<q<0.67515$.
Figure~\ref{fig:zs4} shows the evolution of behavior of the post-measured entropy
$\tilde S(\theta)$ with respect to parameter $q$.
The curve presents a monotonic increase when the parameter $q$ varies from
zero to $q=1/2$. At the point $q=1/2$, as visible in  Fig.~\ref{fig:zs4}(a), a bifurcation of the minimum at $\theta=0$ occurs.
In the range $0.5<q<0.67515$, the curve $\tilde S(\theta)$
reaches, as shown in Figs.~\ref{fig:zs4}(b) and (c), the interior minimum.
So, the region with variable optimal angle $\vartheta$ takes up a part
$0.17515\approx17.5\%$ on the section $[0,1]$ of $q$ axis, and the fidelity between
the states at bound points $q=0.5$ and $q=0.67515$ equals $F=96.8\%$\footnote{
   Note for comparison that experimenters achieve now the values of
	 fidelity $F=99.8(2)\%$ \cite{BSPBG13}, 99.8(1)\% \cite{Guo16}, and 99.998\%
	 \cite{BKABBKK17}.
}.
The position of such a local minimum $\vartheta$ smoothly moves from zero to $\pi/2$ (see again the curves in Figs.~\ref{fig:zs4}(b) and (c) ).

\begin{figure}[b]
\begin{center}
\epsfig{file=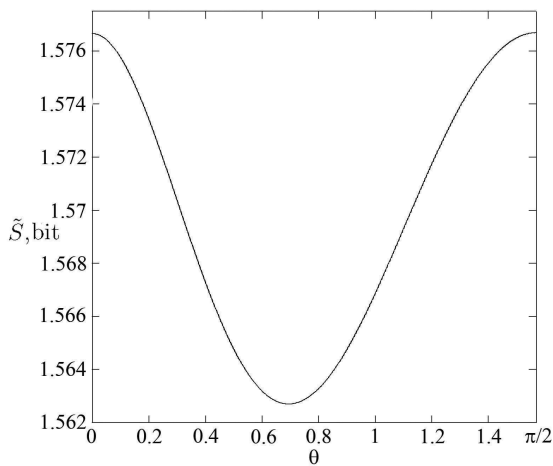,width=7.2cm}
\caption{
High-resolved post-measurement entropy $\tilde S$ vs $\theta$ by $q=0.6155$
}
\label{fig:zs006155m}       
\end{center}
\end{figure}
The interior minimum of post-measurement entropy is best observed when the values of
$\tilde S_0$ and $\tilde S_{\pi/2}$ equal to one another.
This occurs (see Fig.~\ref{fig:zsp01m}) at the point $q=0.61554$ for which
$\tilde S_0=\tilde S_{\pi/2}=1.57667$~bits and hence
$\Delta_{\pi/2}=\Delta_0=q=0.61554$~bit.
The given situation is represented in Fig.~\ref{fig:zs006155m} (cf. with the curve 4
in Fig.~\ref{fig:zs0005-08}).
Here the minimum lies at the angle
$\theta_{min}\equiv\vartheta=0.6955\approx39.8$.
Its depth is 0.01397~bit what yields relative corrections to the post-measurement
entropy and one-way deficit equaled $\delta\tilde S=0.9\%$ and
$\delta{\rm\Delta}=2.3\%$, respectively.

Then, at the value of $q=0.67515$, the system experiences a new hidden sudden
transition -- from the branch, which is characterized by the continuously changing
optimal angle $\vartheta$ in the full interval (from 0 to $\pi/2$), to the branch
$\tilde S_{\pi/2}$ with constant optimal measurement angle equaled $\pi/2$.
From here and up to $q=1$, the curves of post-measured entropy exhibit monotonically
decreasing behavior as illustrated in Fig.~\ref{fig:zs4}(d).

One should emphasize here that the behavior of the minimized one-way quantum deficit
${\rm\Delta}=\min_\theta\Delta(\theta)$ with respect to the argument $q$ is continuous and
smooth.
Nevertheless, the function ${\rm\Delta}(q)$ is a piecewise one,
\begin{equation}
   \label{eq:D2}
   {\rm\Delta}(q)=
	 \cases{
	 \Delta_0,\ &$0\leq q\leq0.5$;\cr
	 \Delta_{\pi/2},\  &$0.67515\leq q\leq1$;\cr
	 \Delta_\vartheta,\ &$q\in(0.5,0.67515)$,
   }
\end{equation}
and therefore presents nonanalyticities at the border points $q=0.5$ and 0.67515 which
manifest themselves in higher derivatives.

\begin{figure}[t]
\begin{center}
\epsfig{file=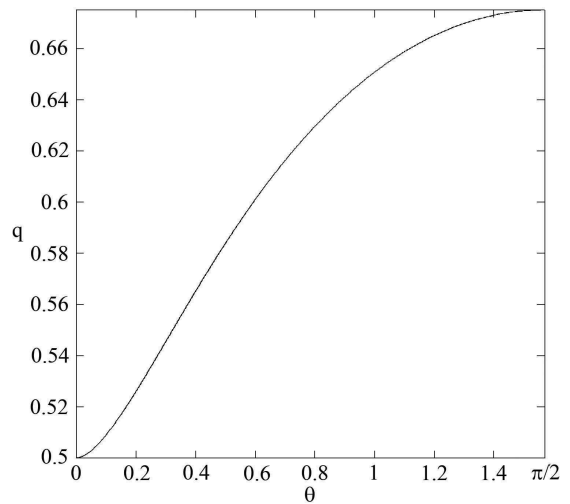,width=7.2cm}
\caption{
Concentration of Bell-diagonal state, $q\in[0.5,0.67515]$, in the mixture
(\ref{eq:rho}) as a function of interior optimal measurement angle $\vartheta$.
}
\label{fig:zq1}       
\end{center}
\end{figure}

The relation of interior optimal measurement on angle $\vartheta$ with
$q\in[0.5,0.67515]$ is shown in Fig.~\ref{fig:zq1}.
The function $q(\vartheta)$ is biunique (one-to-one) and the presented curve allows
to estimate the value of parameter $q$ in the mixed quantum state $\rho_{AB}$.
Hence, this can serve as one of possible applications of quantum correlation in
practice.

\begin{figure*}[t]
\begin{center}
\epsfig{file=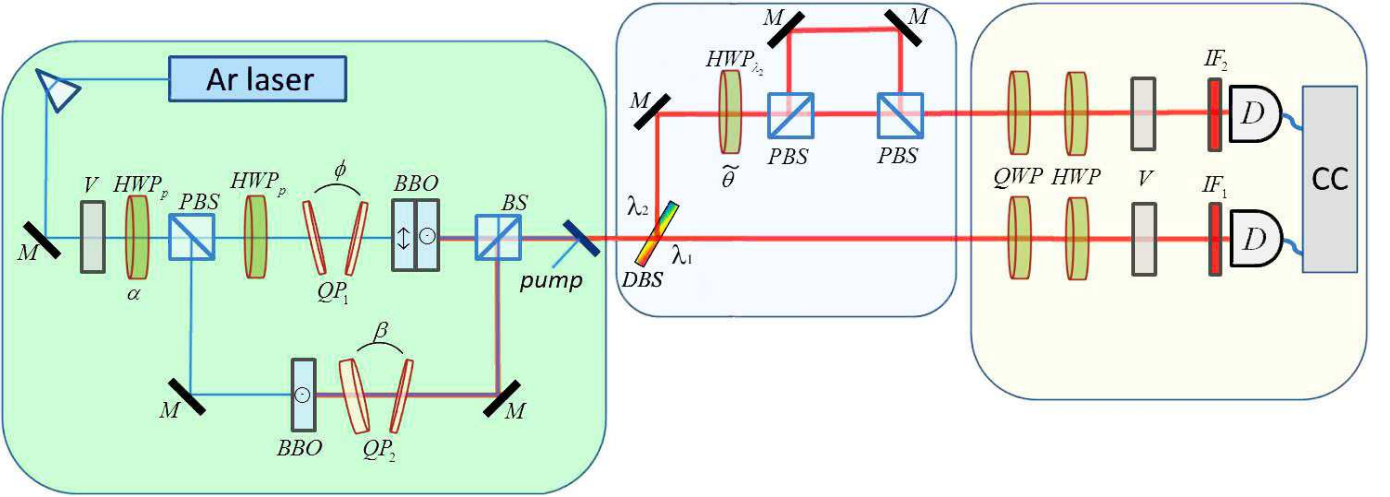,width=18 cm}
\caption{(Color online)
Experimental setup for preparing the initial quantum state and measuring its one-way deficit.
The first (green) block prepares the specific quantum states $\rho_{AB}$ defined in (\ref{eq:rho}),
the second (blue) block performs local measurements
while the third (yellow) block of tomography measures the transformed quantum state $\tilde\rho_{AB}$.
Ar laser: argon laser with wavelength 351~nm,
M: mirror,
V: vertical oriented Glan-Thompson prism,
BBO: nonlinear barium borate crystals of Type-I,
HWP and QWP: half- and quarter-wave plates,
QP: (dichroic) quartz plates,
BS: beamsplitter,
PBS: polarizing beamsplitter,
DBS: dichroic beamsplitter (dichroic mirror),
IF: interference filter,
D: single photon avalanche detectors (SPAD),
CC: coincidence circuit
}
\label{fig:setup}       
\end{center}
\end{figure*}
%
\section{Setup and numerical simulations}
\label{sec:3}

For testing the theoretical approach, described in the previous sections, we propose a design of an experimental setup  allowing to prepare a family of polarization states $\rho_{AB}$(\ref{eq:rho}) and estimate the post-measured entropy $\tilde S$ through the tomographic reconstruction of density matrix.
In particular, we suggest to use the well-known method of polarization state preparation via spontaneous parametric down-conversion process (SPDC) \cite{PRA_our} and quantum state tomography protocol based on tetrahedral symmetry \cite{Rech}, guaranteeing simple realization and high quality of reconstruction.
We intentionally selected commonly used methods to show that the nontrivial behavior of quantum deficit can be observed without requiring specific apparatus.

\subsection{Optical setup scheme}
\label{sec:31}
The setup is schematically depicted in Fig.~\ref{fig:setup}.
A cw argon laser beam at $\lambda=351$ nm passes through a Glan-Thompson prism (V) with vertical orientation, half-wave plate HWP$_{p}$ and polarizing beamsplitter (PBS). The HWP$_{p}$ and PBS serve to control the $q$ parameter in the state $\rho_{AB}$ (\ref{eq:rho}). In the upper arm of non-balanced interferometer the maximum entangled Bell state $|\Phi^+\rangle$ is produced. Two nonlinear type-I BBO crystals, positioned with the planes containing optical axes orthogonal to each other, generate a pair of the basic states $|H_{1}H_{2}\rangle$, $|V_{1}V_{2}\rangle$ via collinear, frequency nondegenerated regime of SPDC. The relative phase $\phi$ between basic states is controlled by two quartz plates QP$_1$, while amplitudes are controlled by the half-wave plate  HWP$_{p}$.

Among th others, and as a practical example, let us suppose that the wavelengths of the collinear downconverted photons are $\lambda_{1}=763$~nm and $\lambda_{2}=650$~nm. In the bottom arm, where the horizontal polarization of the pump is reflected, the second component of the state $\rho_{AB}$ is prepared. We use the technique, suggested in the papers \cite{Jetp_our,PRA_our}, to perform the transformation  $|V_{1}V_{2}\rangle \Rightarrow |H_{1}V_{2}\rangle$. This transformation can be achieved by using dichroic wave plates QP$_2$, which act separately on the photons with different frequencies and
introduce a phase shift of $2\pi$ for a vertically polarized photon at 650~nm, a phase shift of $\pi$ for the conjugated photon at 763~nm. The wave plates are oriented at $45^\circ$ to the vertical direction. Using quartz plates as retardation material it is easy to calculate that the thickness of the wave plate operating the transformation should be equal to 1.585~mm or 3.464~mm. The theoretical estimated fidelity is more than $0.998$, the difference from unit is related to the imperfection of phase transformation for two photons simultaneously.

Since the result of transformation is extremely sensitive to small variations of thickness, we suggest to use the following method to achieve the desired thickness. Two quarts plates with different thicknesses and with orthogonally oriented optical axes correspond to the action of the quartz wave plate with an effective thickness, equal the difference of the thicknesses of two plates. If then one can tilt these wave plates towards each other by a finite angle $\beta$, then the optical thickness of the effective wave plate formed by QP$_2$ will change, and, at a certain value of $\beta$, the desired transformation will be achieved. Ultimately, non-polarizing beamsplitter (BS) mixes the states from the upper and bottom arms of non-balanced interferometer and, as a result, prepares the initial state $\rho_{AB}$.  

The local projective measurements at a variable angle are realized in the (blue) block in Fig.~\ref{fig:setup}. According to Eq.~(\ref{eq:rho_tilde1}) they are implemented only over one of the photon of the pair, so a dichroic beamsplitter (DBS) separates photons with different frequencies into the two spatial modes. In the upper spatial mode a half-wave plate (HWP$_{\lambda_{2}}$) is oriented at the angle $\tilde\theta$ [$\tilde\theta=\theta/4$ is the angle between the input polarization and the wave plate fast axis, $\theta$ is the polar angle in the transformation (\ref{eq:V})] and a couple of polarizing beam splitters (PBS), forming a non-balanced interferometer, perform two orthogonal projections at different angles. To obtain the statistical mixture (\ref{eq:rho_tilde1}) the length difference between the arms of non-balanced interferometer must be larger then the coherence length of the photons with orthogonal polarization, which for the selected wavelength ($\lambda_{2}$) and a full width at half maximum (FWHM) $\delta\lambda=3$~nm of the interference filter IF$_{2}$ equals $\lambda^2/\delta\lambda\approx140$~$\mu$m.

After the non-selective projective measurement the state was sent to the reconstruction block of the setup (the third (yellow) block in Fig.~\ref{fig:setup}). The post-measured quantity of the entropy (\ref{eq:postS}) at varying angle $\tilde\theta$ was numerically calculated through the density matrices of post-measurement state $\tilde\rho_{AB}$ using quantum tomography protocols \cite{PRA_our}. The projective measurements in each arm to perform the quantum state tomography can be realized by means of a polarization filtering system consisting of a sequence of quarter- and half-wave plates, followed by a polarization prism, which transmits vertical polarization ($V$). The detection is operated by taking advantage of silicon single-photon avalanche detectors (SPAD). In the case of independent measurement of two qubits the projective measurements can be chosen arbitrarily. We propose quantum state tomography protocol, where projections on the states possess tetrahedral symmetry \cite{Rech}. There are several works showing that due to the high symmetry such protocol provides a better quality of reconstruction  \cite{Rech,PRL_opt,PRA_opt}. As alternative, adaptive quantum tomography protocols can be used: in fact, despite these require more complex analysis, they guarantee at the same time the highest quality of state reconstruction \cite{PRA_adapt,JETP_adapt}.

\subsection{Numerical simulations}
\label{sec:32}
According to the theory, the region with the variable optimal measurement angle is narrow and therefore requires precise quantum state preparation and reconstruction. It is worth to stress that the standard procedure of the state reconstruction from the likelihood equation associates with finite statistics of the registered outcomes (sample size) of an experiment and therefore takes random values \cite{PRA_our,JETPLBogd}.
For the state $\rho_{AB}$ we can calculate the statistical distribution of fidelity $F$ for the given sample size.

Usually the quantum tomography protocol can be defined by a so-called
instrumental matrix $X$ that has $m$ rows and $s$ columns
\cite{Bogdanov,Bogdanovtwo,krivitsky}, where $s$ is the Hilbert
space dimension and $m$ the number of projections in such space. For
every row, i.e. for every projection, there is a corresponding
amplitude $M_j$.
This matrix equals:
\begin{equation}
M_{j}=X_{jl}c_{l}\qquad (j=1,2,...,m;\quad l =1,2,...,s),\label{eq:ampl}
\end{equation}
where $c_l$ are the expansion coefficients.

The square of the absolute value of the
amplitude defines the intensity of a process, which is the number of
events in one second
\begin{equation}
\lambda_{j}=|M_{j}|^2. \label{eq:prob}
\end{equation}
The number of registered events $k_{j}$ is a random variable
exhibiting Poisson distribution,  $t_{j}$ is the time of exposition of
the selected row of the protocol and $\lambda_j t_j$ the average
value,
\begin{equation}
P(k_j) = \frac{{(\lambda_j t_j )^{k_j}}}{{k_j!}}\exp(-\lambda_j t_j).
\label{eq:poisson}
\end{equation}
The normalization condition for the protocol defines the total expected
number of events $n$ summarized by all rows:
\begin{equation}
\sum\limits_{j = 1}^m {\lambda_j t_j = n}. \label{eq:events}
\end{equation}
Equation (\ref{eq:events}) substitutes the traditional normalization condition for the
density matrices, ${\rm tr}(\rho)=1$.

\begin{figure}[t]
\begin{center}
\epsfig{file=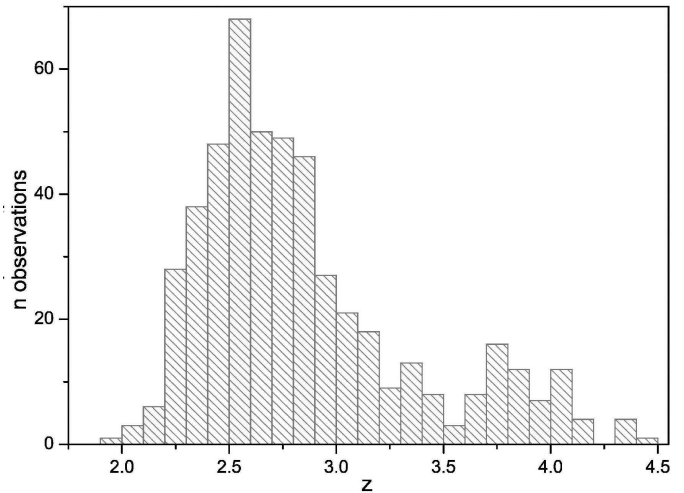,width=8.9cm}
\caption{ The universal statistical fidelity loss histogram distribution for 500 numerical experiments. Sample size of each
experiment is $10^5$}
\label{fig:fidelity}       
\end{center}
\end{figure}

The fidelity achieved now by experimenters has the values $F\geq99.8\% $ for the given optical states (\ref{eq:rho}) \cite{BSPBG13,Guo16,BKABBKK17}, that is why in the numerical experiments testing the universal statistical distribution for fidelity losses we used statistics that guarantees the same fidelity: we numerically generated sets of experimental data and reconstructed $500$ states $\rho_{AB}$ ($q=0.6155$) with sample size around $n=10^5$ for each state, obtained by performing the maximum-likelihood estimation on random variables of the measured counts according to Poissonian statistics. The estimated average fidelity with absolute error is $F=0.998 \pm 0.02$.
The fluctuations of $F$ can be formally taken into account by introducing the so-called loss of fidelity  $1-F$, which is associated only with statistical errors and does not take into account experimental imperfections.We can also introduce the variable $z = - \log_{10}(1-F)$, which is the number of digit 9 in the decimal representation of the parameter $F$ (e.g., $z = 3$ corresponds to $F = 0.999$). Figure~\ref{fig:fidelity} shows the universal statistical fidelity loss histogram distribution for 500 simulated states $\rho_{AB}$. The form of the distribution besides sample size depends on the initial state and the used protocol of quantum state reconstruction. In our simulation the tomography protocol based on the states possessing tetrahedral symmetry had been used
\cite{Rech}.
%
\begin{figure}[t]
\begin{center}
\epsfig{file=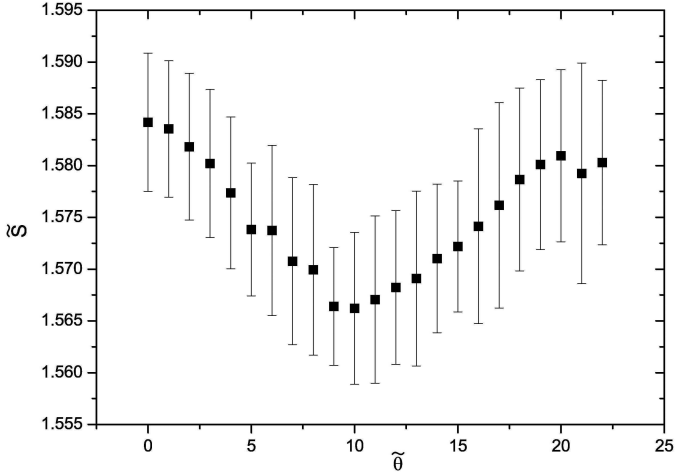,width=8.9cm}
\caption{Post-measurement entropy $\tilde S$ vs $\tilde\theta$ by $q=0.6155$ for 100 numerical experiments. Squares are mean values $\tilde S$, bars are standard deviations. Sample size of each
experiment is $10^5$.
It is clearly seen a minimum of $\tilde S(\tilde\theta)$ near the angle $\tilde\theta=10^\circ$,
that is in good agreement with the theoretical prediction shown in Fig.~\ref{fig:zs006155m}}
\label{fig:entropy}       
\end{center}
\end{figure}
%

For the same sample size we estimated numerically the post-measured entropy for the best observed interior minimum at the point $q=0.6155$. For each angle $\tilde\theta$, which varies from $0^\circ$ to $22^\circ$ (with step $1^\circ$), a set of 100 density matrices of $\tilde\rho_{AB}$ was generated, then eigenvalues were calculated and  finally the mean values together with standard deviations of the post-measured entropy $\tilde S$ were estimated. The dependence $\tilde S$ on $\tilde\theta$ is plotted on Fig.~\ref{fig:entropy}. The graph shows that with a good experimental accuracy, it is possible to observe the region with the variable optimal measurement angle taking advantage of the proposed setup, even if the corrections to the post-measurement entropy values are small (in the order of 0.9\% or less).

\section{Conclusion}
\label{sec:Concl}
In this work we have found the specific quantum state (\ref{eq:rho}) with the record wide region
of variable optimal measurement angle, presented for it the detailed calculations
of entropy after measurement
and hence of one-way deficit, and performed
the numerical simulation of the proposed experiment.

The described experimental setup opens a possibility to construct universal facilities allowing to measure the one-way deficit of any symmetric (up to local unitary
transformation) two-photon quantum states.
Moreover, we have considered in detail the preparation of two-component
mixture of quantum states which, according to the theory, contains a fraction with the
variable optimal measurement angle.

The performed numerical simulations show that the available experimental techniques
allows one to investigate a fine structure of quantum correlation domain including the
region with the variable optimal measurement angle and opens a way to implement the real physical experiment to study the one-way deficit
behavior.

\section*{Acknowledgments}
One of us (M.~Yu.) is grateful to participants of the Seminar of the Quantum
Technologies Center of the
Lomonosov Moscow State University for fruitful discussion.

\section{Author contribution statement}

M.Yu. contributed with theoretical calculations of one-way quantum deficit and wrote the theoretical part of the manuscript, E.M. contributed to the experimental setup design, numerical simulation and wrote the experimental part of the manuscript, M.G. helped with experimental setup design and critical revision of the manuscript.



\begin{thebibliography}{99}

\bibitem{MBCPV12}
K.~Modi, A.~Brodutch, H.~Cable, T.~Paterek, and V.~Vedral,
Rev. Mod. Phys.  \textbf{84}, 1655 (2012)

\bibitem{Str15}
A.~Streltsov,
\textit{Quantum correlations beyond entanglement and their role in quantum information
theory}
(Springer, Berlin 2015)

\bibitem{ABC16}
G.~Adesso, T.~R.~Bromley, and M.~Cianciaruso,
J. Phys. A: Math. Theor. \textbf{49}, 473001 (2016)

\bibitem{FPA17}
\textit{Lectures on general quantum correlations and their applications.}
Edited by F.~F.~Fanchini, D.~O.~Soares-Pinto, and G.~Adesso
(Springer, Berlin 2017)

\bibitem{BDSRSS18}
A.~Bera, T.~Das, D.~Sadhukhan, S.~S.~Roy, A.~Sen(De), and U.~Sen,
Rep. Prog. Phys. \textbf{81}, 024001 (2018)

\bibitem{Z00}
W.~H.~Zurek,
Ann. Phys. (Leipzig) \textbf{9}, 855 (2000)

\bibitem{OZ02}
H.~Ollivier and W.~H.~Zurek,
Phys. Rev. Lett. \textbf{88}, 017901 (2002)

\bibitem{HV01}
L.~Henderson and V.~Vedral,
J. Phys. A: Math. Gen. \textbf{34}, 6899 (2001)

\bibitem{Y11}
M.~A.~Yurishchev,
Phys. Rev. B \textbf{84}, 024418 (2011)

\bibitem{MGY17}
E.~Moreva, M.~Gramegna, and M.~A.~Yurischev,
Adv. Sci. Eng. Med. \textbf{9}, 46 (2017);
arXiv:1612.04589v1 [quant-ph]

\bibitem{OHHH02}
J.~Oppenheim, M.~Horodecki, P.~Horodecki, and R.~Horodecki,
Phys. Rev. Lett. \textbf{89}, 180402 (2002)

\bibitem{HHHHOSS03}
M.~Horodecki, K.~Horodecki, P.~Horodecki, R.~Horodecki, J.~Oppenheim, A.~Sen(De), and U.~Sen,
Phys. Rev. Lett. \textbf{90}, 100402 (2003)

\bibitem{HHHOSSS05}
M.~Horodecki, P.~Horodecki, R.~Horodecki, J.~Oppenheim, A.~Sen(De), U.~Sen, and B.~Synak-Radtke,
Phys. Rev. A \textbf{71}, 062307 (2005)

\bibitem{YF16}
B.-L.~Ye and S.-M.~Fei,
Quantum Inf. Process. {\bf 15}, 279 (2016)

\bibitem{Luo08}
S.~Luo,
Phys. Rev. A {\bf 77}, 042303 (2008)

\bibitem{ARA10}
M.~Ali, A.~R.~P.~Rau, and G.~Alber,
Phys. Rev. A {\bf 81}, 042105 (2010);
Erratum in: Phys. Rev. A {\bf 82}, 069902(E) (2010)

\bibitem{LMXW11}
X.-M.~Lu, J.~Ma, Z.~Xi, and X.~Wang,
Phys. Rev. A {\bf 83}, 012327 (2011)

\bibitem{CZYYO11}
Q.~Chen, C.~Zhang, S.~Yu, X.~X.~Yi, and C.~H.~Oh,
Phys. Rev. A {\bf 84}, 042313 (2011)

\bibitem{LXSW10}
X.-M.~Lu, Z.~Xi, Z.~Sun, and X.~Wang,
Quantum Inf. Comput. {\bf 10}, 0994 (2010)

\bibitem{CRC10}
L.~Ciliberti, R.~Rossignoli, and N.~Canosa,
Phys. Rev. A {\bf 82}, 042316 (2010)

\bibitem{VR12}
S.~Vinjanampathy and A.~R.~P.~Rau,
J. Phys. A: Math. Theor. {\bf 45}, 095303 (2012)

\bibitem{JY16}
N.~Jing and B.~Yu,
J. Phys. A: Math. Theor. {\bf 49}, 385302 (2016)

\bibitem{H13}
Y.~Huang,
Phys. Rev. A {\bf 88}, 014302 (2013)

\bibitem{Y14}
M.~A.~Yurischev,
\textit{Quantum discord for general X and CS states: a piecewise-analytical-numerical
formula},
arXiv:1404.5735v1 [quant-ph]

\bibitem{Y14a}
M.~A.~Yurishchev,
J. Exp. Theor. Phys. {\bf 119}, 828 (2014); arXiv:1503.03316v1~[quant-ph]

\bibitem{Y15}
M.~A.~Yurischev,
Quantum Inf. Process. {\bf 14}, 3399 (2015)

\bibitem{Y17}
M.~A.~Yurischev,
Quantum Inf. Process. {\bf 16}:249 (2017); arXiv:1702.03728v3~[quant-ph]

\bibitem{YWF16}
B.-L.~Ye, Y.-K.~Wang, and S.-M.~Fei,
Int. J. Theor. Phys. {\bf 55}, 2237 (2016)

\bibitem{Y18}
M.~A.~Yurischev,
Quantum Inf. Process. {\bf 17}:6 (2018)

\bibitem{Y18a}
M.~A.~Yurischev,
Quantum Inf. Process. {\bf 18}:124 (2019);
arXiv:1804.03755v2 [quant-ph]

\bibitem{CDSPSS15}
T.~Chanda, T.~Das, D.~Sadhukhan, A.~K.~Pal, A.~Sen(De), and U.~Sen,
Phys. Rev. A {\bf 92}, 062301 (2015)

\bibitem{RSIS09} T.~Ray, H.~K.~Sing, A.~Isaacs, and W.~F.~Smith, ``Infeasibility driven evolutionary algorithm for constrained optimization''. --
In: {\em Constraint-handling in evolutionary optimization}, Edited by E.~Mezura-Montes (Springer, Berlin 2009), pp.~145-165

\bibitem{I12}
M.~G.~Ivanov,
\textit{How to understand the quantum mechanics}
(Regular and Chaotic Dynamics, Moscow-Izhevsk 2012), Sec.~5.3.2 [in Russian]

\bibitem{D30}
P.~A.~M.~Dirac,
\textit{The principles of quantum mechanics}
(Clarendon, Oxford 1930)

\bibitem{vN32}
J.~von~Neumann,
\textit{Mathematische Grundlagen der Quantenmechanik}
(Springer, Berlin 1932)

\bibitem{Z71}
D.~N.~Zubarev,
\textit{Neravnovesnaya statisticheskaya termodinamika}
(Nauka, Moskva 1971) [in Russian]

\bibitem{JKMW01}
D.~F.~V.~James, P.~G.~Kwiat, W.~J.~Munro, and A.~G.~White,
Phys. Rev. A {\bf 64}, 052312 (2001)

\bibitem{BSPBG13}
C.~Benedetti, A.~P.~Shurupov, M.~G.~A.~Paris, G.~Brida, and M.~Genovese,
Phys. Rev. A {\bf 87}, 052136 (2013)

\bibitem{PABJWK04}
N.~A.~Peters, J.~B.~Altepeter, D.~Branning, E.~Jeffrey, T.-C.~Wei, and P.~G.~Kwiat,
Phys. Rev. Lett. {\bf 92}, 133601 (2004);
Erratum in:
Phys. Rev. Lett. {\bf 96}, 159901 (2006)

\bibitem{BMNM04}
M.~Barbieri, F.~De~Martini, G.~Di~Nepi, and P.~Mataloni,
Phys. Rev. Lett. {\bf 92}, 177901 (2004)

\bibitem{APVW07}
A.~Aiello, G.~Puentes, D.~Voigt, and J.~P.~Woerdman,
Phys. Rev. A. {\bf 75}, 062118 (2007)

\bibitem{GGZ11}
F.~Galve, G.~L.~Giorgi, and R.~Zambrini,
Phys. Rev. A {\bf 83}, 012102 (2011);
Erratum in:
Phys. Rev. A {\bf 83}, 069905(E) (2011)

\bibitem{Guo16}
R.~Sun, X.-J.~Ye, J.-S.~Xu, X.-Y.~Xu, J.-S.~Tang, Y.-C.~Wu, J.-L.~Chen, C.-F.~Li, and
G.-C.~Guo,
Phys. Rev. Lett. {\bf 116}, 160404 (2016)

\bibitem{BKABBKK17}
Yu.~I.~Bogdanov, K.~G.~Katamadze, G.~V.~Avosopiants, L.~V.~Belinsky, N.~A.~Bogdanova,
A.~A.~Kalinkin, and S.~P.~Kulik,
Phys. Rev. A {\bf 96}, 063803 (2017)

\bibitem{Jetp_our}
S.~P.~Kulik, G.~A.~Maslennikov, and E.~V.~Moreva,
J. Exp. Theor. Phys. {\bf 105}, 712 (2006)

\bibitem{PRA_our}
Yu.~I.~Bogdanov, E.~V.~Moreva, G.~A.~Maslennikov, R.~F.~Galeev, S.~S.~Straupe, and S.~P.~Kulik,
Phys. Rev. A {\bf 73}, 063810 (2006)

\bibitem{Rech}
J. \v{R}eh\'a\v{c}ek, B.-G. Englert, and D. Kaszlikowski, {Phys. Rev. A} \textbf{70}, 052321 (2004)

\bibitem{PRL_opt}
Yu.~I. Bogdanov, G. Brida, M. Genovese, S.~P. Kulik, E.~V.~Moreva, and A.~P. Shurupov, Phys. Rev. Lett. \textbf{105}, 010404 (2010)

\bibitem{PRA_opt}
Yu.~I. Bogdanov, G.~Brida, I.~D.~Bukeev, M.~Genovese, K.~S.~Kravtsov, S.~P.~Kulik, E.~V.~Moreva, A.~A.~Soloviev, and A.~P.~Shurupov, Phys. Rev. A \textbf{84}, 042108 (2011)

\bibitem{PRA_adapt}
G. I. Struchalin, I. A. Pogorelov, S. S. Straupe, K.~S.~Kravtsov, I. V. Radchenko, and S. P. Kulik,
Phys. Rev. A \textbf{93}, 012103 (2016)

\bibitem{JETP_adapt}
S. S. Straupe, Pis'ma v ZhETF {\bf 104},  540 (2016)

\bibitem{JETPLBogd} Yu.~I. Bogdanov, J. Exp. Theor. Phys. \textbf{108}, 928 (2009)

\bibitem{Bogdanov} Yu.~I.~Bogdanov, M.~V.~Chekhova, L.~A.~Krivitsky, S.~P.~Kulik, A.~N.~Penin, L.~C.~Kwek, A.~A.~Zhukov, C.~H.~Oh, and M.~K.~Tey, Phys. Rev. A \textbf{70}, 042303 (2004)

\bibitem{Bogdanovtwo} Yu. I. Bogdanov,
{\em Statistical inverse problem: root approach}, arXiv:0312042v1 [quant-ph]

\bibitem{krivitsky} Yu. I. Bogdanov, L. A. Krivitsky, and S. P. Kulik, {JETP Lett.}
\textbf{78}, 352 (2003)

\end{thebibliography}
\end{document}